\documentclass[twocolumn]{aastex62}

\graphicspath{{./}{figures/}}

\accepted{\today}
\submitjournal{ApJL}

\shorttitle{The deepest \textit{Chandra} view of RBS 797}
\shortauthors{Ubertosi et al.}

\begin{document}

\title{The deepest \textit{Chandra} view of RBS 797: evidence for two pairs of equidistant X-ray cavities}

\correspondingauthor{F. Ubertosi}
\email{francesco.ubertosi2@unibo.it}

\author[0000-0001-5338-4472]{F. Ubertosi}
\affil{Dipartimento di Fisica e Astronomia, Università di Bologna, via Gobetti 93/2, I-40129 Bologna, Italy}
\affil{INAF, Osservatorio di Astrofisica e Scienza dello Spazio, via P. Gobetti 93/3, 40129 Bologna, Italy}

\author[0000-0002-0843-3009]{M. Gitti}
\affil{Dipartimento di Fisica e Astronomia, Università di Bologna, via Gobetti 93/2, I-40129 Bologna, Italy}
\affil{Istituto Nazionale di Astrofisica (INAF) - Istituto di Radioastronomia, via Gobetti 101, I-40129 Bologna, Italy}

\author[0000-0001-9807-8479]{F. Brighenti}
\affil{Dipartimento di Fisica e Astronomia, Università di Bologna, via Gobetti 93/2, I-40129 Bologna, Italy}
\affil{University of California Observatories/Lick Observatory, Department of Astronomy and Astrophysics, University of California, Santa Cruz, CA 95064, USA}

\author[0000-0003-4195-8613]{G. Brunetti}
\affil{Istituto Nazionale di Astrofisica (INAF) - Istituto di Radioastronomia, via Gobetti 101, I-40129 Bologna, Italy}

\author[0000-0001-5226-8349]{M. McDonald}
\affil{Kavli Institute for Astrophysics and Space Research, Massachusetts Institute of Technology, Cambridge, MA 02139, USA}

\author[0000-0003-0297-4493]{P. Nulsen}
\affil{Chandra X-ray centre, Smithsonian Astrophysical Observatory, 60 Garden Street, Cambridge, MA 02143, USA}

\author[0000-0001-7224-0650]{B. McNamara}
\affil{Department of Physics and Astronomy, University of Waterloo, 200 University Avenue West, Waterloo, ON N2L 3G1, Canada}
\affil{Perimeter Institute for Theoretical Physics, Waterloo, ON N2L 2Y5, Canada}
\author[0000-0002-3984-4337]{S. Randall}
\affil{Center for Astrophysics, Harvard \& Smithsonian, 60 Garden St., Cambridge, MA 02138, USA}

\author[0000-0002-9478-1682]{W. Forman}
\affil{Smithsonian Astrophysical Observatory, Harvard-Smithsonian Center for Astrophysics, 60 Garden St., Cambridge, MA 02138, USA}

\author[0000-0002-2808-0853]{M. Donahue}
\affil{Department of Physics and Astronomy, Michigan State University, East Lansing, MI 48824, USA}

\author[0000-0003-1581-0092]{A. Ignesti}
\affil{INAF - Astronomical Observatory of Padova, vicolo dell'Osservatorio 5, IT-35122 Padova, Italy}
\author[0000-0003-2754-9258]{M. Gaspari}
\affil{INAF, Osservatorio di Astrofisica e Scienza dello Spazio, via P. Gobetti 93/3, 40129 Bologna, Italy}
\affil{Department of Astrophysical Sciences, Princeton University, 4 Ivy Lane, Princeton, NJ 08544, USA}

\author[0000-0003-4117-8617]{S. Ettori}
\affil{INAF, Osservatorio di Astrofisica e Scienza dello Spazio, via P. Gobetti 93/3, 40129 Bologna, Italy}
\affil{INFN, Sezione di Bologna, viale Berti Pichat 6/2, 40127 Bologna, Italy}

\author[0000-0003-0312-6285]{L. Feretti}
\affil{Istituto Nazionale di Astrofisica (INAF) - Istituto di Radioastronomia, via Gobetti 101, I-40129 Bologna, Italy}

\author{E. L. Blanton}
\affil{Institute for Astrophysical Research and Astronomy Department, Boston University, 725 Commonwealth Avenue, Boston, MA 02215, USA}
\author{C. Jones}
\affil{Harvard-Smithsonian Center for Astrophysics, 60 Garden Street, Cambridge, MA 02138, USA}

\author[0000-0002-2238-2105]{M. Calzadilla}
\affil{Kavli Institute for Astrophysics and Space Research, Massachusetts Institute of Technology, Cambridge, MA 02139, USA}




\begin{abstract}
We present the first results of a deep \textit{Chandra} observation of the galaxy cluster RBS 797, whose previous X-ray studies revealed two pronounced X-ray cavities in the east-west (E-W) direction. Follow-up VLA radio observations of the central active galactic nucleus (AGN) uncovered different jet and lobe orientations, with radio lobes filling the E-W cavities and perpendicular jets showing emission in the north-south (N-S) direction over the same scale ($\approx$30 kpc). With the new $\sim$427 ks total exposure, we report the detection of two additional, symmetric X-ray cavities in the N-S direction at nearly the same radial distance as the E-W ones. The newly discovered N-S cavities are associated with the radio emission detected at 1.4 GHz and 4.8 GHz in archival VLA data, making RBS 797 the first galaxy cluster found to have four equidistant, centrally-symmetric, radio-filled cavities. We derive the dynamical and radiative ages of the four cavities from X-ray and radio data, respectively, finding that the two outbursts are approximately coeval, with an age difference of $\lessapprox$10 Myr between the E-W and N-S cavities. We discuss two scenarios for the origin of the two perpendicular, equidistant cavity systems: either the presence of a binary AGN which is excavating coeval pairs of cavities in perpendicular directions, or a fast ($<$10 Myr) jet reorientation event which produced subsequent, misaligned outbursts. 
\end{abstract}

\keywords{galaxies: clusters: individual (RBS 797) - galaxies: clusters: intracluster medium - X-rays: galaxies: clusters - radio continuum: galaxies}

\section{Introduction} \label{sec:intro}
\noindent The impact of supermassive black hole (SMBH) activity in brightest cluster galaxies (BCGs) on the surrounding intracluster medium (ICM) is best traced by surface brightness depressions, ripples and filaments in the X-ray images of galaxy clusters. In particular, the discovery of X-ray cavities thought to be excavated by radio lobes launched from the central active galactic nucleus (AGN) has built momentum in the study of AGN feeding and feedback mechanisms in cool core galaxy clusters and groups (e.g., \citealt{1993MNRAS.264L..25B,2000A&A...356..788C,2000ApJ...534L.135M,2000MNRAS.318L..65F,2004ApJ...607..800B,2005ApJ...625L...9N,2007ApJ...659.1153W,2010ApJ...714..758G,2011ApJ...732...13G,2015ApJ...805..112R,2015ApJ...811..111M}). These nearly circular depressions, which are typically found in pairs, can observationally be associated with bright rims of cold gas (e.g., \citealt{2014MNRAS.442.3192V}). Moreover, cocoon shocks driven by the outburst and encompassing the cavities have been detected with deep observations in some clusters (e.g., \citealt{2014MNRAS.442.3192V,2019MNRAS.484.3376L}). 
\\ Images of radio-filled cavities not only provide instant snapshots of the cluster conditions, but can also unveil the history of the AGN-ICM interaction: the detection of multiple pairs of X-ray cavities, each excavated every few tens of Myr, traces the radio-X-ray interplay and the AGN duty cycle over time (e.g., \citealt{2005MNRAS.363..891F,2005MNRAS.364.1343D,2015ApJ...805...35H}). In this context, obtaining precise estimates of the outburst ages is crucial to address how the activity cycles of the central engine are coupled with the ICM thermodynamic state (e.g., for reviews \citealt{2007ARA&A..45..117M,2012NJPh...14e5023M,2012AdAst2012E...6G,2012ARA&A..50..455F,2016NewAR..75....1S,2020NatAs...4...10G,2021Univ....7..142E}). Exquisite examples of multiple X-ray cavities have been found by \textit{Chandra}, with the various pairs either aligned along a common axis (e.g., NGC 5813, \citealt{2015ApJ...805..112R}; MS 0735, \citealt{Biava2021}; Hydra A, \citealt{2007ApJ...659.1153W}) or misaligned, following either jet reorientation (e.g., Cygnus A, \citealt{2012A&A...545L...3C}) or cavity sideways motion in the ICM (e.g., Perseus, \citealt{2010ApJ...713L..74F}). 
\\ In light of the link between cavities and radio lobes, the relative position of multiple cavity pairs can trace the direction of the jets that excavated each depression. It has been proposed that radio galaxies experiencing changes in jet direction could be harbouring binary SMBHs (e.g., \citealt{2002Sci...297.1310M}): the companion of the primary, accreting AGN can either induce precession of the jet axis (e.g., \citealt{2018JApA...39....8R}) or, if active itself, produce a secondary pair of jets (e.g., 3C75 \citealt{2006A&A...453..433H}). Jet reorientation in radio active BCGs is expected to affect the ICM conditions: large reorientation angles or rapid jet direction changes could prompt an isotropic distribution of heating \citep{2018A&A...617A..58C,2020AAS...23514404L}.  
\newline
\\The galaxy cluster RBS 797 (RA 09:47:12.76, DEC +76:23:13.74, z=0.354) is a perfect case study for investigating the above topics: an early \textit{Chandra} observation (Cycle 2, 13.3 ks, \citealt{2001A&A...376L..27S}) unveiled two deep X-ray cavities in the east-west (E-W) direction surrounded by bright rims, clearly confirmed with a longer observation (Cycle 8, 38 ks, \citealt{2011ApJ...732...71C,2012ApJ...753...47D}). Multi-frequency radio observations with the VLA disclosed radio emission with different orientations: at 1.4 GHz and 4.8 GHz, radio lobes originating from the central AGN fill the X-ray cavities (at $\sim$30 kpc from the center, \citealt{2006A&A...448..853G,2013A&A...557L..14G}), and two additional lobe-like extensions up to $\sim$27 kpc are present in the north-south (N-S) direction (i.e. with a $\sim$90$^{\circ}$ misalignment w.r.t. the E-W cavities, \citealt{2006A&A...448..853G,2013A&A...557L..14G}). 
\\ At sub-arcsec resolution, two pairs of kpc-scale jets oriented in the E-W and N-S direction have been discovered at 4.8 GHz \citep{2013A&A...557L..14G}. Moreover, a short observation of the radio core with the European VLBI Network (EVN) at 5 GHz uncovered two compact components separated by $\sim$77 pc \citep{2013A&A...557L..14G}. Combining these findings, \citet{2013A&A...557L..14G} suggested that either the central SMBH has experienced reorientation events, producing subsequent outbursts in perpendicular directions, or the BCG hosts a dual AGN, with the two SMBHs launching jets in perpendicular directions (coeval outbursts). 
\\While further radio analysis of the nuclear region is required to confirm \textit{whether} the BCG hosts a dual AGN (dedicated multi-frequency VLBI observations - EVN, eMerlin and VLBA - are currently being analysed, PI: Gitti), RBS 797 lacks measurements of the various outbursts' ages, and a direct assessment of the N-S outburst imprints on the surrounding ICM, which are crucial to understand \textit{how} the nuclear activity is coupling with the cluster environment.
These key questions have driven the request for deeper \textit{Chandra} data, to understand if the N-S radio emission is associated with shallow X-ray cavities (missed by the existing images), and to probe whether the multiple, misaligned AGN outburst are very efficiently heating the cluster core.
\\In this article we present the first results of the analysis of the new \textit{Chandra Large Program} observation of RBS 797 (Cycle 21 LP proposal, 420 ks, PI: Gitti), focusing on the X-ray cavities and the AGN outburst history. A thorough investigation of the whole cool core region will be presented in a forthcoming paper (Ubertosi et al., in prep.). \\We assume a $\Lambda$CDM cosmology with H$_{0}$=70 km s$^{-1}$ Mpc$^{-1}$, $\Omega_{m}$=0.3 and $\Omega_{\Lambda}$=0.7, which gives a 1$''$=4.9 kpc conversion at z=0.354. Uncertainties are reported at $1\sigma$. The radio spectral index $\alpha$ is defined as $S_{\nu}\propto\nu^{-\alpha}$.
\section{The Data} \label{sec:data}
\subsection{X-ray - Chandra} \label{subsec:xray}
\noindent The new \textit{Chandra} data for RBS 797 have been acquired in VFAINT mode during Cycle 21 (ObsIDs: 22636, 22637, 22638, 22931, 22932, 22933, 22934, 22935, 23332, 24631, 24632, 24852, 24865), for a total exposure time of 409 ks. Adding the previous observations (ObsIDs 2202, 7902) the overall exposure reaches $\sim$458 ks. Data have been reprocessed using CIAO-4.13 and CALDB-4.9.0. The \texttt{wavdetect} tool was used to obtain a list of point sources, which were masked during the analysis. After excluding the time intervals during background flares the cleaned exposure is $\sim$427 ks. Background files were obtained from blanksky-event files, reprojected to match the ObsIDs and normalized to the 9-12 keV count rate of the observations.
\begin{figure*}[h]
	\centering
	\gridline{\fig{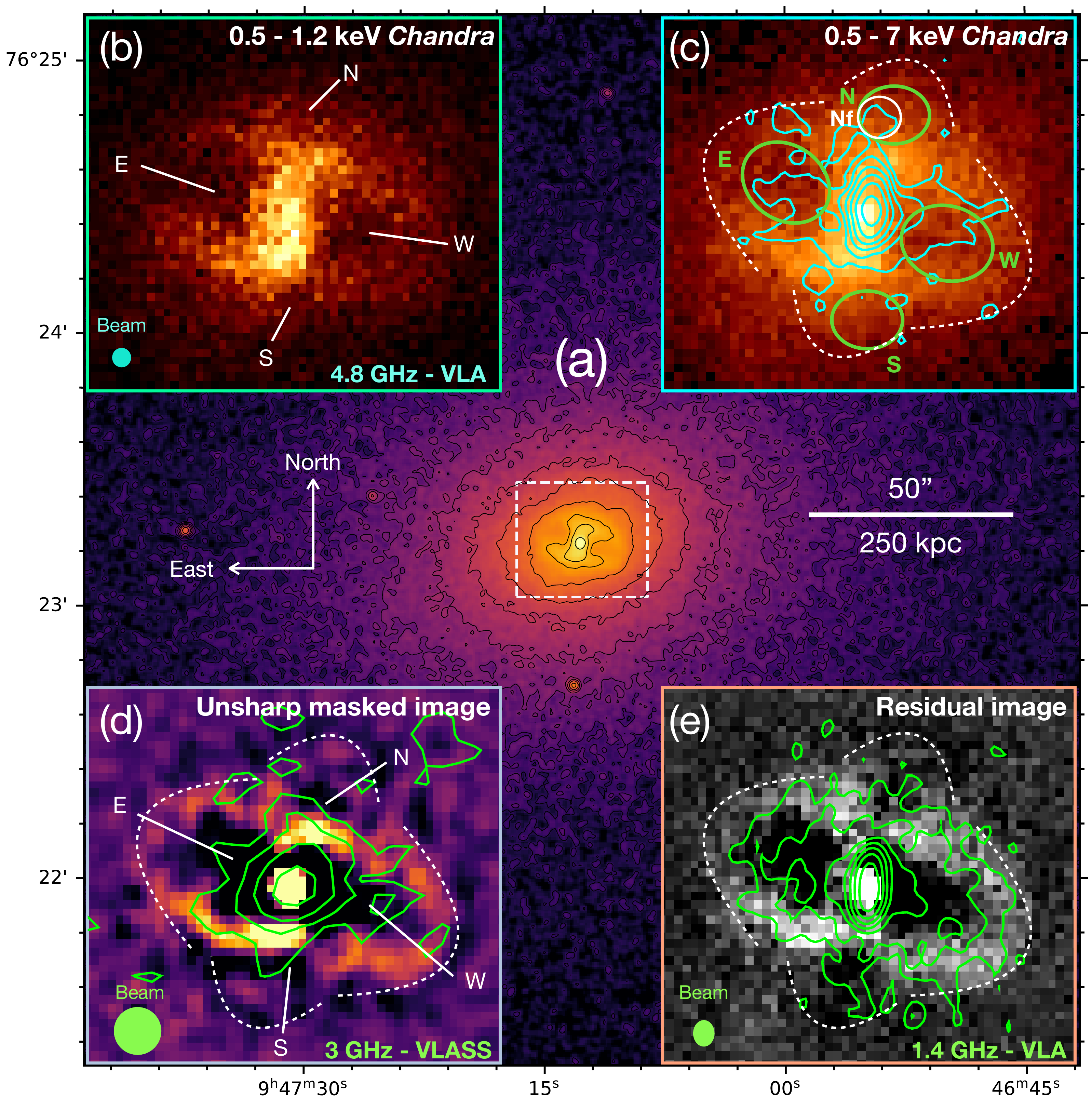}{1\textwidth}{}}
	\caption{Exposure-corrected, background-subtracted mosaiced \textit{Chandra} images of RBS 797. (\textit{a}) 0.5-7 keV image, with a white dashed box indicating the region covered by the zoom-ins. Black contours are spaced by a factor of 2, with the highest being 4$\times$10$^{-6}$ cts s$^{-1}$ cm$^{-2}$. (\textit{b}) 0.5-1.2 keV image of the core; white labels indicate the position of the X-ray cavities. (\textit{c}) 0.5-7 keV image of the core, with ellipses showing the shape of the E-W-N (Nf)-S cavities, and white dashed arcs encompassing the rims (see text for details). Cyan contours at 4.8 GHz (rms$=$0.01 mJy/beam, $\sim$1.3$''$ resolution, \citealt{2013A&A...557L..14G}) show the morphology of the radio galaxy. (\textit{d}) Unsharp masked image of RBS 797, obtained by subtracting a 3$''$ smoothed image from a 1$''$smoothed one, with white dashed arcs as in panel \textit{c}. Green contours at 3 GHz are from the VLASS (contours spaced by a factor of 2 starting from 3$\times$rms$=$0.1 mJy/beam, 2.5$''$ resolution). (\textit{e}) Double $\beta$-model residual image of RBS 797, with white dashed arcs as in panel \textit{c}. Green contours at 1.4 GHz (rms$=$0.02 mJy/beam, $\sim$1.5$''$ resolution) are from \citet{2006A&A...448..853G}.}
	\label{fig:cavities}
\end{figure*}
\subsubsection{Morphological analysis}\label{subsubsec:morpho}
\noindent Mosaiced exposure-corrected, background subtracted images of RBS 797 have been built using two energy bands (see Fig. \ref{fig:cavities}\textit{a}, \ref{fig:cavities}\textit{b} and \ref{fig:cavities}\textit{c}). A soft band (0.5 - 1.2 keV) image of the cluster core was used to accentuate the cavities and the surrounding rims. A total band (0.5-7 keV) image was used to derive unsharp masked and residual images, and to measure the significance of the depressions discussed in Sect.\ref{sec:result} (see Fig. \ref{fig:sig}).
\\ To produce an unsharp masked image we tested several smoothing scale combinations (between 1$''$-5$''$). The unsharp masked image in Fig. \ref{fig:cavities}\textit{d} has been obtained by subtracting two images, smoothed with a Gaussian of 1$''$ and 3$''$ kernel size, respectively; this choice best emphasizes structures in the cluster core. In particular, we verified that the X-ray cavities discussed in Sect.\ref{sec:result} are recovered regardless of the specific choice, therefore we are confident that they are not spurious features.
\noindent As an alternative and complementary method to accentuate morphological features, we modeled the emission of the ICM with two elliptical $\beta$-models on SHERPA and then subtracted it off the image (e.g., \citealt{2015ApJ...811..111M,2019ApJ...887L..17C}). 
The orientation and eccentricity of the two models were left free to vary, while the centers were fixed to the X-ray peak. The resulting residual image is shown in Fig. \ref{fig:cavities}\textit{e}. 

\subsection{Radio - VLA} \label{subsec:radio}
\noindent To investigate the radio plasma-ICM interaction, we relied on the VLA data at 1.4 GHz (configuration A, B and C, see \citealt{2006A&A...448..853G,2012ApJ...753...47D}), and at 4.8 GHz (configuration A and B, see \citealt{2013A&A...557L..14G}), which best trace the radio emission from the BCG. \\ Our main aim was to measure for the first time the spectral index and radiative age of each AGN outburst. To do so, we re-analysed the archival data (using standard reduction procedures in AIPS-\texttt{31DEC20}). The total intensity maps we produced are consistent with those already published, therefore the radio contours used in this work are from \citet{2006A&A...448..853G,2013A&A...557L..14G}. To compute the spectral index we produced maps at 1.4 GHz and 4.8 GHz with the \texttt{IMAGR} task setting uniform weighting (ROBUST=-5), matching uv-range (2.8-176 k$\lambda$) and clean beam size ($1.6''\times1.1''$). Such maps were used to measure fluxes and compute spectral indices in different regions (see Sect. \ref{subsec:rages} and Fig. \ref{fig:spixprofiles}). \\RBS 797 was also observed during the VLASS survey\footnote{\url{https://science.nrao.edu/vlass}.} \citep{2020PASP..132c5001L} at 3 GHz. In order to characterize the radio source at this frequency so far unobserved, we generated radio contours from the survey maps (see Fig. \ref{fig:cavities}\textit{d}).

\section{Results} \label{sec:result}
\subsection{The detection of two additional X-ray cavities} \label{subsec:detect}
\noindent The mosaiced 0.5 - 1.2 keV \textit{Chandra} image of Fig. \ref{fig:cavities}\textit{b} captures the details of the AGN - ICM interaction in RBS 797: the already known E and W X-ray cavities are perfectly visible as deep, elliptical depressions surrounded by bright rims and located at $\sim$5.5$''$ ($\sim$27 kpc) from the BCG. As also noted by \citet{2011ApJ...732...71C}, the rims are not symmetric, being brighter in an S-shaped region starting south of the E cavity and ending north of the W cavity. 
\\The E-W cavities in RBS 797 have previously been associated with the radio lobes of the BCG \citep{2006A&A...448..853G,2011ApJ...732...71C,2012ApJ...753...47D}. As mentioned in the introduction, \citet{2013A&A...557L..14G} discovered N-S extensions of the radio galaxy. Therefore, to investigate the imprints of AGN activity on the ICM in the orthogonal direction to the E-W cavities, we searched for X-ray counterparts to the observed N-S radio lobes. Interestingly, the deeper \textit{Chandra} exposure unveils two additional depressions north and south of the nucleus. The two depressions are almost perpendicular to the larger E-W cavities, and nearly at the same projected distance from the AGN ($\sim$5.5$''\sim$27 kpc). 
\\The unsharp and residual images (Fig. \ref{fig:cavities}\textit{d} and \ref{fig:cavities}\textit{e}) emphasize the N-S holes, which appear to be surrounded by faint rims: the northern rim is unambiguous in the residual and unsharp images, and can be identified in the 0.5-7 keV image as a strip of brighter pixels above the N feature. The southern one corresponds to a bright blob in the unsharp masked image below the S depression. 
\begin{figure*}
	\centering
	\gridline{\fig{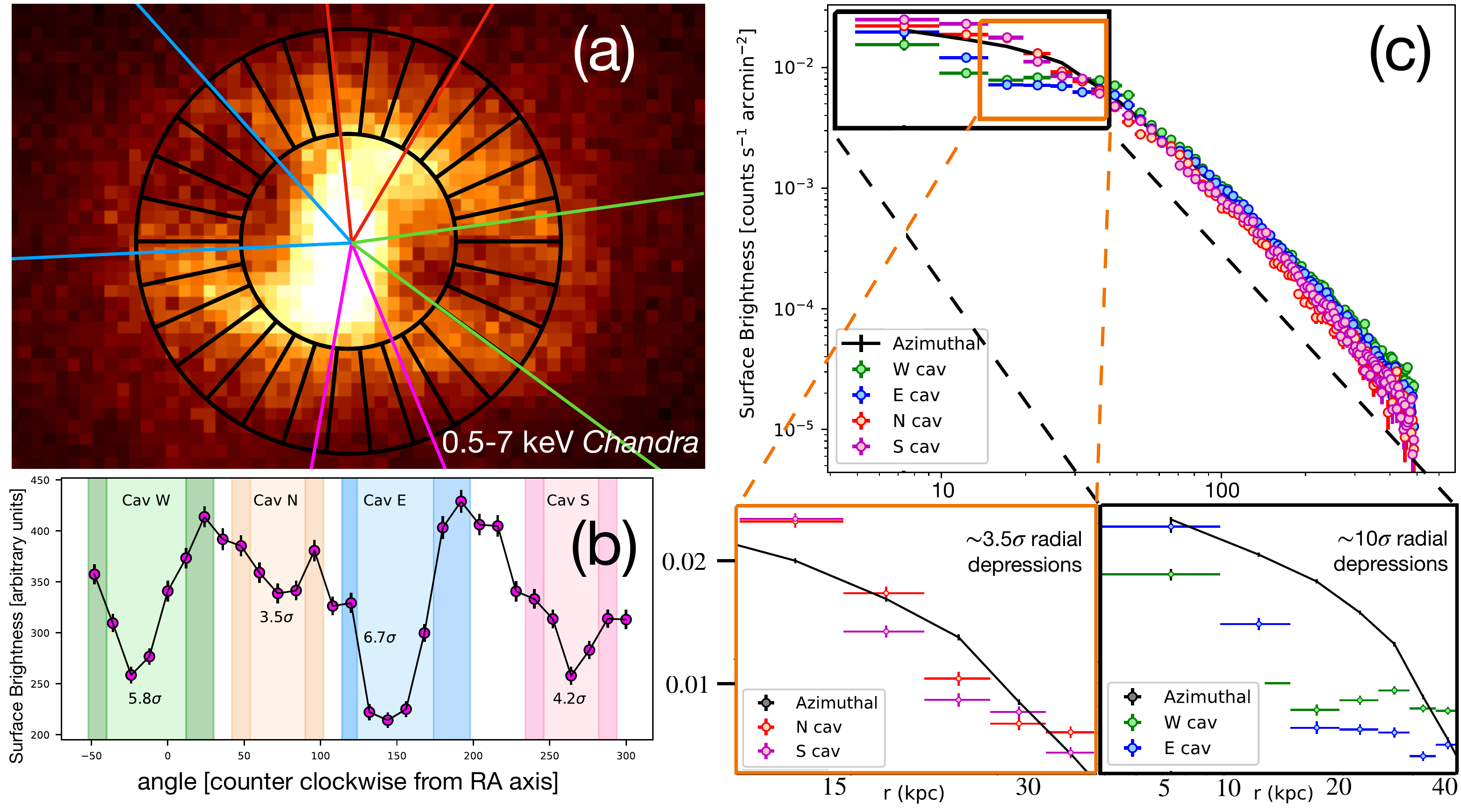}{\textwidth}{}}
	\caption{(\textit{a}) 0.5 - 7 keV band \textit{Chandra} image of the core: black sectors used for the azimuthal analysis and colored sectors along each cavity used for the radial analysis are superimposed. (\textit{b}) Azimuthal variation of surface brightness measured in 30 sectors extending 4$''$-8$''$ from the center. Light colored rectangles correspond to each cavity, while darker strips indicate reference regions used to measure the significance. (\textit{c}) Background-subtracted 0.5–7 keV surface brightness profile (black line) of the cluster (cavities excluded) in 1$''$ bins, compared to profiles along the E-W cavity regions (right zoom-in) and N-S cavity regions (left zoom-in).}
	\label{fig:sig}
\end{figure*}
\\ We used ellipses to describe each cavity, the semi-axes of which have been chosen by comparing the unsharp, residual and original image (see Tab. \ref{tab:cavage}). For the N depression, we considered an alternative configuration (Nf), whose size represents the fraction of feature N filled by radio emission both at 1.4 GHz and 4.8 GHz (see Fig. \ref{fig:cavities}).
\\Putative depressions N and S of the nucleus were originally identified by \citet{2011ApJ...732...71C}, who also noted the coincidence of radio emission with the X-ray structures. However, a clear detection of these structures was lacking due to the short exposure.
Our deep Chandra observation allows us to securely detect the new N-S depressions, thus identifying  RBS 797 as the first galaxy cluster for which two symmetric X-ray cavity pairs coincident with resolved radio emission are found at the same projected distance from the center. This peculiar geometry challenges classical methods to estimate the cavity significance (i.e. surface brightness deficits w.r.t. the azimuthal mean, see e.g., \citealt{2015ApJ...805...35H}). As the whole azimuth is disturbed either by part of the rims or by the cavities themselves, it is nontrivial to define a reference surface brightness. In particular, the already known E-W depressions are deep, thus the shallower N-S depressions would lie above the azimuthal mean. In turn, the significance of the N-S cavities would be based on a surrounding surface brightness which is influenced by the presence of the E-W depressions. 
\\ To circumvent this obstacle, we considered the annulus from 4$''$ to 8$''$, divided into 30 equal sectors (Fig. \ref{fig:sig}\textit{a}) and then relied on the comparison between the surface brightness measured in the sectors covering each cavity, $S_{\text{c}}$ (with uncertainty $E_{\text{c}}$), and that measured in the immediate surroundings, $S_{\text{s}}$ (with uncertainty $E_{\text{s}}$). The results are shown in Fig. \ref{fig:sig}\textit{b}, where the the lighter and darker strips correspond to $S_{\text{c}}$ and $S_{\text{s}}$, respectively. The surface brightness decrement of each structure (D) and the significance at which the decrement is recovered, D over the error in D, were computed as: 
\begin{equation}
     \text{D} = 1 - \frac{S_{\text{c}}}{S_{\text{s}}} 
\end{equation}
\begin{equation}
    \text{significance} = \frac{\text{D}}{(1-\text{D})\sqrt{\Big(\frac{E_{\text{c}}}{S_{\text{c}}}\Big)^{2} + \Big(\frac{E_{\text{s}}}{S_{\text{s}}}\Big)^{2}}}
\end{equation}
\noindent The N-S cavities also represent radial deficits w.r.t. the azimuthally averaged radial surface brightness profile (extracted after masking the cavities, e.g., \citealt{2012ApJ...753...47D}; see Fig \ref{fig:sig}\textit{c}). Overall, in the E-W cavities the net counts drop by about D = 30\% with a significance ranging from $\sim$6-10$\sigma$ depending on the method adopted, while in the N-S structures the net counts drop by about D = 10$\%$ at 3.5 - 4.2$\sigma$ (based on the analysis shown in Fig. \ref{fig:sig}\textit{b}, \ref{fig:sig}\textit{c}). We made various tests by varying the choice for the sectors, also considering elliptical annuli, and always found consistent results. Additionally, we verified that N-S dips can be recovered by extracting a surface brightness profile from two linear projections (one to the north and one to the south of the center, respectively) along a straight cut parallel to the axis of the E-W cavities (a similar method has been previously used to test the existence of small cavities in MS 0735, see \citealt{2014MNRAS.442.3192V}). These alternative methods are shown in Appendix \ref{appendixA}. For comparison, the other dark structures visible in the unsharp masked image have typically 5\% fewer counts at about 1$\sigma$ confidence only. Additionally, we followed the method outlined in \citet{2020ApJ...897...57M} to measure cavity signal-to-noise ratios (SNR) w.r.t. the double $\beta$-model. For each cavity region, the counts in the image of Fig. \ref{fig:cavities}\textit{c}, $N_{\text{I}}$, were compared with the counts in the double $\beta$-model residual image of Fig. \ref{fig:cavities}\textit{e}, $N_{\text{M}}$. The deficit of the structures is defined as:
\begin{equation}
    \text{Deficit} = 1 - \frac{N_{\text{I}}}{N_{\text{M}}},
\end{equation}
and the SNR for the deficit is (see Eq. 9 in \citealt{2020ApJ...897...57M}):
\begin{equation}
    \text{SNR} = \frac{\mid N_{\text{I}} - N_{\text{M}}\mid}{\sqrt{\mid N_{\text{I}} - N_{\text{M}}\mid + 2N_{\text{M}}}}
\end{equation}
\noindent Using the above equations, the N-S cavities represent deficits of about 10\% at a SNR of 5, while the deeper E-W cavities have deficits of about 25\% at a SNR of 17. These estimates are consistent with the deficits based on the observed surface brightness, thus strengthening our results.
\\Such considerations suggest that the N-S depressions represent an additional system of X-ray cavities, created by the central AGN. The suggested presence of rims surrounding the N-S cavities further strengthens this interpretation. The new cavity detection is strongly supported by the morphology of the central radio source: the 1.4 GHz contours overlaid on Fig. \ref{fig:cavities}\textit{e} show radio lobes coincident with the E-W cavities, and reveal significant extensions partly filling the N-S depressions. The 4.8 GHz emission (Fig. \ref{fig:cavities}\textbf{\textit{c}}) extends towards the E-W cavities over the same scales as the 1.4 GHz contours. Furthermore, a third structure extends northward, terminating in a lobe coincident with the N cavity (partially filling it within the Nf sub-region). The 3 GHz VLASS contours (2.5$''$ resolution) overlaid on Fig. \ref{fig:cavities}\textit{d} confirm the cross-like morphology of the AGN, with four radio extensions headed towards the perpendicular cavity pairs. Thus, the co-spatiality of significant radio emission with the four depressions is key to demonstrating that the N-S holes are real X-ray cavities and not artificial dips produced by the ``$\mathbf{\infty}$ - shaped'' morphology of the bright rims. 
\\ We note that RBS 797 might resemble e.g., 2A0035+096 \citep{2009MNRAS.396.1449S} or 4C+00.58 \citep{2010ApJ...717L..37H}, in which multiple X-ray cavities have been found near the cluster core; in these clusters, however, the association of radio emission from the central AGN with the cavities is not straightforward. Furthermore, we observe the similarity of RBS 797 to Cygnus-A, in which additional cavities filled with radio emission and perpendicular to the main cavity system have been discovered (see \citealt{2012A&A...545L...3C}). The peculiarity of RBS 797 is that, for the first time, multiple, perpendicular systems of radio-filled X-ray cavities are found at the same projected distance from the BCG.

\subsection{Cavity ages from X-ray data} \label{subsec:Xages}
\noindent Measuring the age of the cavities is crucial to investigate the activity cycle of the central AGN. \citet{2004ApJ...607..800B} originally discussed three methods to measure cavity ages from X-ray data, typically in agreement within a factor of 2: 
\begin{itemize}
    \item The sonic time $t_{\text{cs}} = D_{\text{AGN}}/c_{\text{s}}$, which is the time required for the cavity to reach its distance from the BCG, $D_{\text{AGN}}$, by moving at the ICM sound speed $c_{\text{s}}=\sqrt{\gamma kT/\mu m_{\text{p}}}$; here $\gamma=5/3$ is the adiabatic index, $kT$ is the ICM temperature at $D_{\text{AGN}}$, $\mu=0.61$ is the molecular weight, and $m_{\text{p}}$ is the proton mass.
    \item The buoyancy time $t_{\text{buo}} = D_{\text{AGN}}/\sqrt{2gV/0.75S}$, where $g$ is the gravitational acceleration at the cavity position, while $V$ and $S$ are the cavity volume and cross section to the flow, respectively.
    \item The refill time $t_{\text{ref}} = 2\sqrt{r_{\text{eff}}/g}$, where $r_{\text{eff}}=\sqrt{ab}$ is the effective radius of an ellipsoid of semi-axes $a$ and $b$, and $g$ is the gravitational acceleration. 
\end{itemize}
\noindent We derived the ages of the E, W, N (Nf), S cavities in RBS 797 following the above methods, and assuming that the cavities all lie in the plane of the sky. The cavities were treated as prolate ellipsoids to compute the volume; a systematic uncertainty of $10\%$ was included for distances and semi-axes (e.g., \citealt{2003ApJ...587..619S}). Since the sound speed depends on the ambient temperature, and $g$ can be derived from the ICM pressure gradient and density by assuming hydrostatic equilibrium\footnote{Hydrostatic equilibrium is uncertain in systems with strong AGN feedback signatures (e.g., \citealt{2011MNRAS.411..349G}); however, since the cavities are found at the same projected distance from the center, even a difference in $g$ of a factor of e.g. 2 would make the buoyancy and refill ages of the two cavity pairs scale by the same factor ($\sqrt{g}\sim$1.4). As we are interested in relative ages (see Sect. \ref{sec:disc}), deviations from hydrostatic equilibrium are not very important.} \citep{2006MNRAS.368..518V}, we built radial profiles of thermodynamic quantities by extracting spectra from concentric circular annuli (cavities excluded). 
\\The spacing of these annuli was chosen to reach a SNR$>70$ for each spectrum (or about 10000 counts in each region), that resulted in 23 radial bins. Spectra were fitted with a \texttt{projct$\ast$tbabs$\ast$apec} model in XSPEC-v.12.10. We defer the complete presentation of the detailed spectral analysis of RBS 797 to a forthcoming paper (Ubertosi et al., in prep.); in this work we employ the ICM temperature $kT$=4.04$^{+0.29}_{-0.26}$ keV and electron density $n_{\text{e}}$=6.52$^{+0.22}_{-0.21}\times10^{-2}$ cm$^{-3}$ measured at the distance of the cavities from the center (in the radial bin encompassing the cavities, between 5$''$-6$''$), which are key to derive the sound speed $c_{\text{s}}$=1008$\pm$54 km s$^{-1}$ and acceleration $g$=5.4$\pm$0.6$\times10^{-8}$ cm s$^{-2}$. The above temperature and electron density are consistent with those reported in \citet{2011ApJ...732...71C,2012ApJ...753...47D}. We note that measuring local values of $c_{\text{s}}$ in four sectors (one for each cavity) produced negligible differences ($\approx6\%$) w.r.t. the azimuthally averaged value reported above. 
\begin{table*}
	\centering
	\caption{Properties of the four X-ray cavities in RBS 797.}
	\label{tab:cavage}
		\begin{tabular}{lccccccccc}
			\hline
			Cavity & D$_{\text{AGN}}$ & a & b & r$_{\text{eff}}$ & t$_{\text{cs}}$ & t$_{\text{buo}}$ & t$_{\text{ref}}$ & t$_{\text{exp,cs}}$& t$_{\text{rad}}$ \\
			& (kpc ['']) & (kpc ['']) & (kpc ['']) & (kpc ['']) & (Myr) & (Myr) & (Myr) & (Myr) & (Myr) \\
			\hline
			
			\rule{0pt}{4ex} E & 27.9 (5.7) & 15.7 (3.2) & 12.7 (2.6) & 14.1 (2.9) & 26.5$\pm$1.7 & 26.8$\pm$2.4& 56.7$\pm$4.2& 13.4$\pm$1.6& 37$\pm$21 \\
			
			\hline
			
			\rule{0pt}{4ex} W & 25.5 (5.2) & 15.7 (3.2) & 12.7 (2.6) & 14.1 (2.9) & 23.8$\pm$1.5 & 24.1$\pm$2.2 & 56.7$\pm$4.2& 13.4$\pm$1.6& 36$\pm$18\\
			
			\hline
			
			\rule{0pt}{4ex} N & 29.4 (6.0) & 9.8 (2.0) & 7.4 (1.5) & 8.5 (1.7) & 28.8$\pm$1.9 & 36.1$\pm$3.2 & 43.9$\pm$3.3 & 8.1$\pm$1.0 & $<$28\\
			
			\rule{0pt}{4ex} Nf & 28.4 (5.8) & 5.9 (1.2) & 5.4 (1.1) & 5.6 (1.15) & 27.0$\pm$1.7 & 46.2$\pm$4.3& 36.0$\pm$2.7 & 5.4$\pm$0.6& 14$\pm$10\\
			
			\hline
			
			\rule{0pt}{4ex} S & 27.0 (5.5) & 9.8 (2.0) & 7.4 (1.5) & 8.5 (1.7) & 26.0$\pm$1.7 & 32.6$\pm$3.4& 43.9$\pm$3.3 & 8.1$\pm$1.0& $<$38  \\
		
			\hline
		\end{tabular}
		\tablecomments{(1) Cavity name (Nf is the portion of cavity N filled with radio emission at 1.4 GHz and 4.8 GHz, see Fig. \ref{fig:cavities}); (2) Projected distance from the center of the radio galaxy; (3) semi-major axis; (4) semi-minor axis; (5) effective radius ($\sqrt{ab}$); (6) sound crossing time; (7) buoyancy time; (8) refill time; (9) expansion time; (10) radiative age, obtained from the spectral index within each cavity. See Sect. \ref{sec:result} for details.}
\end{table*}
\\Additionally, we propose a further method to constrain the ages: we computed the time required for each cavity to reach its observed size by expanding at the sound speed, $t_{\text{exp}}=r_{\text{eff}}/c_{\text{s}}$. This method assumes either that the cavity has been excavated directly at the position where it is observed, or that it has expanded and risen at the same speed ($c_{\text{s}}$), reaching a distance from the center comparable to its radius.
\\ For the E-W cavities, both $t_{\text{cs}}$ and $t_{\text{buo}}$ suggest an average outburst age of 25$\pm$2 Myr (consistent with \citealt{2012ApJ...753...47D}). The refill time and the sound expansion time are a factor of $\sim$2 higher and lower, respectively. The N-Nf-S cavities have a $t_{\text{cs}}$ comparable to that of the E-W cavities (which is expected considering the dependencies on $D_{\text{AGN}}$ and $c_{\text{s}}$). On the other hand, the buoyancy and refill time suggest average ages of $\approx$40 Myr. The sound expansion times set the lowest age estimates of $\approx$5-8 Myr. The complete results of the four methods are reported in Tab. \ref{tab:cavage}.
\\ We caution that the uncertainties on ages reported above and in Tab. \ref{tab:cavage} include only statistical uncertainties (which are small given the depth of the \textit{Chandra} exposure), since systematic uncertainties related to the assumptions of each method are not known. Projection effects represent an additional source of uncertainty for the position and 3D structure of the cavities, as we do not
know the inclination of either the E-W or N-S cavities with respect to
the plane of the sky. For instance, if the four cavities were buoyantly rising in the cluster atmosphere, from the definition of $t_{\text{buo}}$ we should expect coeval outbursts to be found at the same distance from the center and to have similar sizes. In this context, we note that the moderate discrepancy of buoyancy ages between the two cavity pairs (a factor of $\approx$1.5) is mostly related to the different ratio between the observed size and distance from the center. On the one hand, the difference in ratio is likely to be affected by projection effects, so that the observed distance and size of the cavities might be underestimated. On the other hand, the different $D_{\text{AGN}}/r_{\text{eff}}$ could hint at AGN of different jet kinetic power (see e.g., \citealt{2008ApJ...687..173D}) which have inflated equidistant holes of varying sizes. 
\\ \noindent The above considerations suggest that it is not straightforward to draw conclusions on the cavity dynamics based on the age computed with a single method. Therefore, we decided not to prefer any method over the others, and to interpret the range in ages returned by the four estimates as an indication of our uncertainty for the true outburst ages (see also Sect. \ref{sec:disc}).
 \begin{figure*}
	\centering
	\gridline{\fig{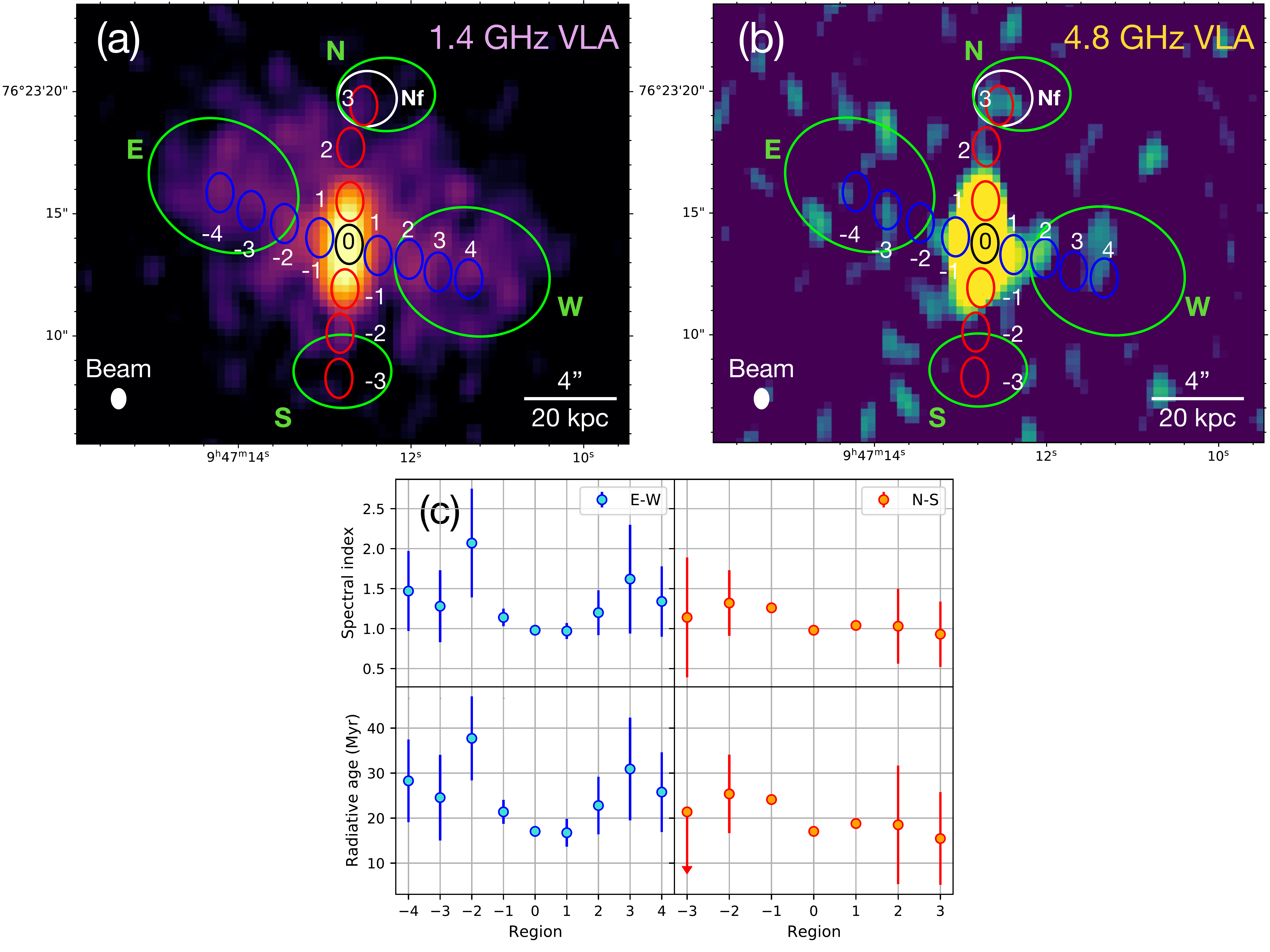}{1\textwidth}{}}
	\caption{Radio maps (panels \textit{a} and \textit{b}) used to compute the spectral index and radiative age (panel \textit{c}). The radio maps are derived with uniform weighting, matching uv-range (2.8-176 k$\lambda$), and clean beam size 1.6$''$×1.1$''$. (\textit{a}) 1.4 GHz VLA image, obtained from the combined A-B-C archival observations. The rms is 0.02 mJy/beam. (\textit{b}) 1.4 GHz VLA image, obtained from the combined A-B archival observations. The rms is 0.01 mJy/beam. Overlaid on the maps are the cavity regions (green and white) and the beam-sized ellipses used to sample the spectral index and radiative ages (panel \textit{c}) in the E-W (blue) and N-S (red) directions.}
	\label{fig:spixprofiles}
\end{figure*}
\subsection{Radiative ages from VLA data} \label{subsec:rages}
\noindent Since the radio lobes are thought to have excavated the cavities, a comparison between the dynamical ages derived from the X-ray analysis and the radiative age of the synchrotron-emitting plasma should return consistent results (see e.g., \citealt{2020MNRAS.496.1471K,Biava2021}). By considering only the synchrotron and Inverse Compton losses of relativistic electrons, the shape of the radio spectrum can be analysed to derive radiative ages, once the magnetic field strength is known. Unfortunately, the available radio data of RBS 797 do not allow detailed spectral fitting of a sychrotron ageing model (KP, JP or Tribble, \citealt{1962AZh....39..393K,1970ranp.book.....P,1973A&A....26..423J,1993MNRAS.261...57T}; for recent applications see e.g., \citealt{2013MNRAS.435.3353H}), which would require flux measurements at three different frequencies at least, while maintaining the resolution required to image the four lobes ($\leq$2$''$).
\\ As an approximation, it is possible to measure spectral indices $\alpha$ between $\nu_{1}=$1.4 GHz and $\nu_{2}=$4.8 GHz, and obtain radiative ages using the following equation (e.g., \citealt{2014NJPh...16d5001E,2019A&A...631A.173B}): 
\begin{equation}
    t_{\text{rad}}[\text{Myr}] = \frac{1590\sqrt{B}}{(B^{2} + B_{CMB}^{2})\sqrt{1+z}}\sqrt{\frac{(\alpha-\Gamma)\text{ln}(\frac{\nu_{2}}{\nu_{1}})}{\nu_{2} - \nu_{1}}}
    \label{eq:trad}
\end{equation}
\noindent where $B$ and $B_{CMB}=3.25(1+z)^{2}$ are the source and the equivalent Cosmic Microwave Background magnetic fields, respectively ($\mu$G), and $\Gamma$ is the injection index. We checked that this method, while approximate (it assumes $\nu\gtrapprox\nu_{b}$, where $\nu_{b}$ is the break frequency of the synchrotron spectrum, which is a reasonable assumption) provides consistent results with those of spectral modeling (see also \citealt{2019A&A...631A.173B}).
We used the minimum energy loss field $B=B_{CMB}/\sqrt{3}=3.4\mu$G (see \citealt{2017SciA....3E1634D}), that maximizes the radiative lifetime of the emitting particles (we also verified that assuming equipartition returns consistent results). The injection index $\Gamma$ was set to 0.7 (typical values can range from 0.5 to 0.9, see \citealt{Biava2021} and references therein). Since, with the available uv-sampling at 4.8 GHz, we recover flux from the extended lobes only at the $\sim$2$\sigma$ level, our results suffer from relatively large uncertainties (up to a factor of 2). 
\\ As a note of caution, we observe that using $B=3.4\mu$G and neglecting adiabatic expansion losses of the relativistic electrons results in possibly overestimated radiative ages; adopting steeper values of $\Gamma$ would result in even lower ages. On the other hand, the spectrum of the integrated flux from the cavity regions could also be described with a continuous injection model \citep{1970ranp.book.....P}, which would result in higher radiative ages than those predicted using Eq. \ref{eq:trad}. These arguments are particularly important for the Nf cavity: assuming $\Gamma\sim$ 0.8-0.9 (close to the observed spectral index) would result in its radiative age being close to zero. However, a detailed comparison between different models and injection indices goes beyond the accuracy that can be reached with the available radio data, which only provide approximate radiative ages.
\begin{figure*}
	\centering
	\gridline{\fig{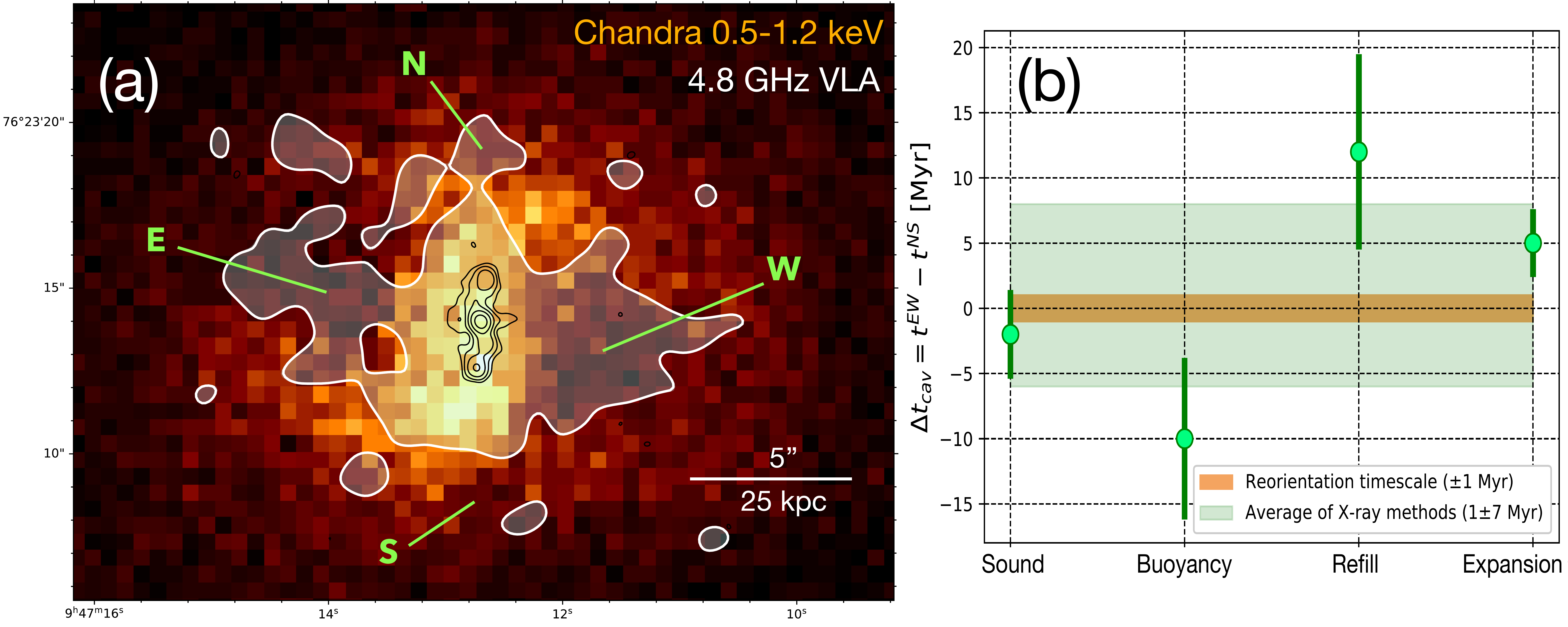}{1\textwidth}{}}
	\caption{(\textit{a}) 0.5-1.2 keV image of the core of RBS 797. Black contours at 4.8 GHz (rms$=$0.01 mJy/beam, $\sim$0.5$''$ resolution, \citealt{2013A&A...557L..14G}) show the morphology of the four, perpendicular, inner jets of the radio galaxy. The filled white contour at 4.8 GHz is the largest contour (at 3$\times$rms) shown in Fig. \ref{fig:cavities}\textbf{\textit{c}}. Green labels indicate the four X-ray cavities. (\textit{b}) The age difference of the E-W and N-S cavities for the different methods is shown and compared to the theoretical reorientation timescale ($\pm$1 Myr, orange area; see Sect. \ref{sec:disc}).}
	\label{fig:last}
\end{figure*}
\\ Fig. \ref{fig:spixprofiles} shows a sampling of the spectral indices with beam-sized ellipses ($1.6''\times1.1''$) in the E-W (blue ellipses in panel \textit{a}) and N-S (red ellipses) directions. The spectral index profiles (panel \textit{c}) show no strong steepening towards the outer regions: the innermost region has a spectral index of $\sim$1 (consistent with \citealt{2006A&A...448..853G}), which is steeper than typical core emission; however, this may be explained by the relatively large beam ($1.6''\times1.1''$), which does not resolve the core-jet components. The $t_{rad}$ profiles show that from the center to the outer regions the E-W radiative ages range between 15-35 Myr, while the N-S ones range between 15-25 Myr. 
\\ Additionally, we computed the integrated flux from the cavity regions to obtain the radiative ages within each feature (last column in Tab. \ref{tab:cavage}). The E-W ages are consistent with both buoyancy, sound speed and refill times. For the N-S cavities, we only have upper limits; the Nf cavity has $t_{\text{rad}}\approx$14 Myr, which is lower than the timescales suggested by buoyancy and refill ages, and is closer to the sound speed age. 

\section{Discussion} \label{sec:disc}
\noindent In this section we discuss the implications of our results on the estimated ages of the two outbursts. As mentioned in the introduction, \citet{2006A&A...448..853G,2013A&A...557L..14G} discovered four perpendicular jets in the innermost $\sim15$ kpc of the radio galaxy (see Fig. \ref{fig:last}\textit{a}), and tentatively identified two compact radio core components. To explain this peculiar radio morphology, \citet{2013A&A...557L..14G} originally proposed that the perpendicular outbursts were produced either following a reorientation event of the AGN jets (subsequent scenario), or by the contemporaneous activity of two SMBHs in the BCG of RBS 797 (coeval scenario). In the following, we focus on whether the four X-ray cavities we discovered in the follow-up \textit{Chandra} observation can help in discriminating between the subsequent and coeval scenarios. 
\\ Overall, our joint X-ray-radio analysis has unveiled that the two perpendicular outbursts have comparable ages: the $t_{\text{exp}}$ and $t_{\text{rad}}$, considered as lower and upper limits for the true age, respectively, constrain the four cavities to be $\sim$10-50 Myr old. Specifically, Fig. \ref{fig:last}\textit{b} shows the age difference between the two outbursts, computed as $\Delta t_{cav} = t^{EW}-t^{NS}$. Here $t^{EW}$ and $t^{NS}$ represent, for each method, the average age of each cavity pair. It is interesting to note that the X-ray methods do not agree on which cavity system is older: the suggested time interval is at most $\pm$10 Myr, with the average of the X-ray methods lying approximately at zero (1$\pm$7 Myr, green area in Fig. \ref{fig:last}\textit{b}).
\\Similarly, the radio analysis returns a slightly different age between the E-W and N-S outbursts: 
from the $t_{\text{rad}}$ of the cavities (Tab. \ref{tab:cavage}), we obtain $t^{EW}\lesssim$36 Myr and $t^{NS}\lessapprox$28-38 Myr; consistently, the beam-sampled ages in Fig. \ref{fig:spixprofiles}\textit{c} between E-W and N-S are similar, indicating a difference of $\sim$10 Myr, at most.
\\Altogether, if the four cavities are the consequence of a reorientation event, the above timescales suggest that the AGN axis changed rather quickly, specifically in $\lesssim$10 Myr. It is noteworthy that the range of ages in RBS 797 is short when compared to the mean outburst interval of known galaxy clusters with multiple generations of cavities ($\sim$100 Myr in the sample of \citealt{2004ApJ...607..800B}; $\sim$20 Myr for Perseus and Virgo, \citealt{2013ApJ...768...11B}; $\sim$50-60 Myr for MS0735, \citealt{2014MNRAS.442.3192V,Biava2021}). Additionally, a radio spectral index study for the X-shaped radio galaxy A3670 (X-shaped radio galaxies are believed to have experienced reorientation events, see e.g., \citealt{liu2004}) found a difference of $\sim$20 Myr between the lobes and the wings \citep{2019A&A...631A.173B}, which is larger than our estimate of $\lesssim$10 Myr. Therefore, considering the above timescales on which misaligned outbursts are typically observed, it is possible that the two outbursts in RBS 797 might have been inflated during the nearly contemporaneous activity of two SMBHs. 
\\ We note that with the available data we cannot definitely exclude that the two outbursts are not coeval on timescales shorter than our uncertainties. In RBS 797, the mean age difference between the E-W and N-S cavities is $\Delta t_{cav}$=1$\pm$7 Myr (from the average of the X-ray methods, whereas the radiative ages provide only an upper limit of $\lesssim$10 Myr). The timescales for jet reorientation are uncertain, but theoretical works suggest that the reorientation could typically happen in $\Delta t_{\text{reo}}\approx1$ Myr (orange area in Fig. \ref{fig:last}\textit{b}), or longer if the jets are switched off during the event (see e.g., \citealt{2002MNRAS.330..609D,liu2004,2007MNRAS.374.1085L}). A rapid change in the AGN feeding can also cause such a fast jet reorientation; e.g., chaotic cold accretion has been shown to flicker very rapidly in less than a Myr, with a sudden change in angular momentum \citep{2017MNRAS.466..677G}. While it is possible that reorientation happened on timescales shorter than $\sim$1 Myr, the reoriented jets would take $\approx10$ Myr to propagate in the ICM (assuming a jet advance speed of 0.01 c, e.g., \citealt{2021MNRAS.503.5948M}) and inflate radio lobes. The derived AGN outburst timescales in RBS 797 thus seem at the boundary between a fast jet reorientation and coeval, binary SMBHs activity; our analysis indicates that the latter might be possible, given the $\Delta t_{cav}\lesssim$10 Myr being sensibly smaller than literature values for misaligned cavity systems.
\\ Therefore, our results are still consistent with a coeval scenario, that is the presence of a binary system of \textit{two active} SMBHs in the core of RBS 797, possibly coincident with the two compact components discovered by \citet{2013A&A...557L..14G}. In particular, the detection in the \textit{Chandra} images of RBS 797 of four perpendicular, equidistant X-ray cavities with similar ages is a tantalizing indication for the presence of two AGNs. 

\section{Conclusion} \label{sec:conc}
\noindent Overall, the new, deep Chandra data revealed that RBS 797 is a beautiful example of how strongly the AGN activity can impact the ICM conditions. Our results can be summarized as follows:
\begin{itemize}
    \item With the deepest to date \textit{Chandra} observation of RBS 797 (427 ks) we unveiled the existence of two additional X-ray cavities (in the N-S direction) at the same distance from the center as the previously known E-W cavities ($\sim$27 kpc). Using archival, multi-frequency VLA data, we highlighted the co-spatiality of the four cavities with the radio lobes of the central radio galaxy. Thus, we find RBS 797 to be the first system in which four equidistant, radio filled X-ray cavities are symmetrically found in perpendicular directions. 
    \item We computed the ages of the cavities by means of both the X-ray and radio data: the four cavities have similar ages, being approximately 10-50 Myr old, and with an age difference between the two outbursts of at most $\sim$10 Myr. 
    \item Considering the properties of the four symmetrical X-ray cavities and the inferred timescales, we argue that the scenario in which a binary AGN is powering the two perpendicular outbursts might be preferred, although a rapid jet reorientation of a single AGN cannot be excluded. With the upcoming deep VLBI study of the radio core it may be possible to provide a final answer on the origin of the two perpendicular outbursts discovered with \textit{Chandra}. 
\end{itemize}

\acknowledgments
\noindent We thank the anonymous referee for constructive and useful comments. Support for this work was provided to MM and MC by the National Aeronautics and Space Administration through Chandra Award Number GO0-21114A issued by the Chandra X-ray Center, which is operated by the Smithsonian Astrophysical Observatory for and on behalf of the National Aeronautics Space Administration under contract NAS8-03060. Additional support was provided to MM and MC from the Space Telescope Science Institute, which is operated by the Association of Universities for Research in Astronomy, Inc., under NASA contract NAS 5–26555. This support is specifically associated with the program HST-GO-16001. AI acknowledges the Italian PRIN-Miur 2017 (PI A. Cimatti). We acknowledge financial contribution from the agreement ASI-INAF n.2017-14-H.0 (PI Moretti). WF acknowledges support from the Smithsonian Institution, the Chandra High Resolution Camera Project through NASA contract NAS8-03060, and NASA Grants 80NSSC19K0116, GO1-22132X, and GO9-20109X. SWR is supported by the Chandra X-ray Center through NASA contract NAS8-03060, and the Smithsonian Institution.
%

\vspace{5mm}
\facilities{CXO, VLA.}


\software{astropy \citep{2013A&A...558A..33A,2018AJ....156..123A},  
          APLpy \citep{2012ascl.soft08017R}, Numpy \citep{2011CSE....13b..22V,2020Natur.585..357H}, Scipy \citep{jones2001scipy}, CIAO \citep{2006SPIE.6270E..1VF}, XSPEC \citep{1996ASPC..101...17A}, AIPS \citep{van1996aips}. 
          }


\appendix
\section{Alternative methods to detect the N-S cavities}
\label{appendixA}
\begin{figure*}
	\centering
	\gridline{\fig{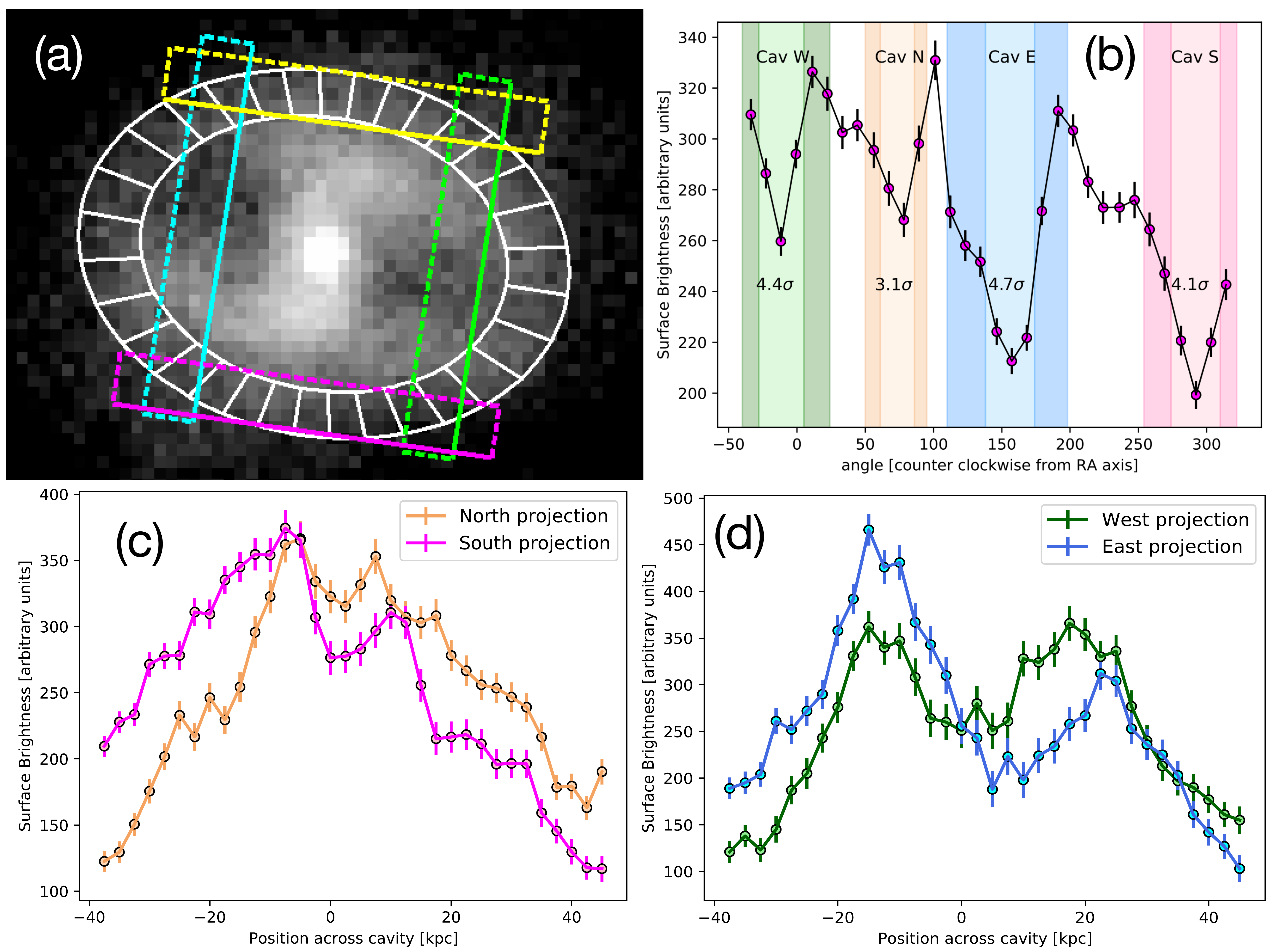}{\textwidth}{}}
	\caption{(\textit{a}) 0.5 - 7 keV band \textit{Chandra} image of the core: white sectors used for the azimuthal analysis of panel \textit{b} and projection regions used for the profiles of panels \textit{c} and \textit{d} are superimposed. (\textit{b}) Azimuthal variation of surface brightness measured in the sectors of panel \textit{a}. (\textit{c}) Plot of surface brightness along the linear projections shown in panel \textit{a} for the N-S cavities. (\textit{d}) Plot of surface brightness along the linear projections shown in panel \textit{a} for the E-W cavities.}
	\label{fig:projections}
\end{figure*}

\noindent This Appendix shows alternative methods used to confirm the presence of N-S cavities in the ICM. \paragraph{Elliptical sector} Figure \ref{fig:projections}\textit{a} shows a different choice for the sector used to study azimuthal variations of surface brightness. We used an elliptical region that encompasses the centers of the N-S cavities, crosses the terminal part of the E-W cavities and avoids the bright rims of the E-W cavities. The ellipticity and orientations were chosen to follow the elliptical structure of the cavity region, while the center matches that of Fig. \ref{fig:sig}\textit{a}. Our starting hypothesis was that the elliptical symmetry could allow to define a proper reference surface brightness, in order to check whether the N-S depressions lie below an azimuthally symmetric \textit{mean} surface brightness. Figure \ref{fig:projections}\textit{b} shows the resulting azimuthal study, with the cavity regions and reference regions coloured as in Fig. \ref{fig:sig}. We note that by using the same method defined in Sect. \ref{sec:result} to measure significance, thus selecting the immediate surroundings of each cavity as references, we obtain similar results to those that rely on circular symmetry. Moreover, we note that the elliptical sector does not provide azimuthal symmetry for the surface brightness. The surface brightness distribution is asymmetric: the whole N-W side of the cluster is brighter than the S-E side. This supports our hypothesis that the mean, being influenced by the large amount of structures in the ICM around the center (as discussed in Sect. \ref{sec:result}), would not be good choice of reference surface brightness.
\paragraph{Projection regions} Following the method outlined in \citet{2014MNRAS.442.3192V}, we selected linear projections (one to the north and one to the south of the center, respectively) along a straight cut parallel to the axis of the E-W cavities (see Fig. \ref{fig:projections}\textit{a}). The N-S cavities correspond to 10\% - 15\% deficits w.r.t. the surrounding small rims (see Fig. \ref{fig:projections}\textit{c}). For comparison, Fig. \ref{fig:projections}\textit{d} shows the same method applied to the deeper E-W cavities (20\% - 30\% deficits w.r.t the surrounding bright rims). These results are consistent with those presented in Fig. \ref{fig:sig}.




\bibliographystyle{aasjournal.bst} 
\bibliography{rbs797.bib}






\end{document}